# The FetalSenseM V2: a wearable and wireless, multi-wavelength NIRS device for continuous monitoring of placental oxygenation and metabolism

**L. Giannoni[a], F. Lange[a], M. Talati[b], U. Hakim[a], D. Gopal[a], A. Artemiou[a], N. Ranaei-Zamani[c], Olivia Newth[c], Sarah Johnson[c], Maeve McGaughey[c], S. Mitra[c], and I. Tachtsidis[a]**

[a] Department of Medical Physics and Biomedical Engineering, University College London, London, UK
[b] School of Physics and Astronomy, University of Glasgow, Glasgow, UK
[c] Neonatology, EGA Institute for Women's Health, University College London, London, UK.
[*]l.giannoni@ucl.ac.uk

**Abstract** We present here the FetalSenseM V2 (FSM V2), a wireless and wearable, multi-wavelength, NIRS device, based on LED illumination at six NIR wavelengths (782, 800, 818, 830, 848 and 888 nm). It utilises two sources located on the flexible wings of the sensor body, and three photodiodes placed symmetrically in its centre at 1 cm from each other. The configuration provides two sets of three source-to-detector distances (SDS) at 3, 4 and 5 cm. FSM V2 allows estimation of dynamic, relative changes in placental concentrations of $HbO_2$, HHb and oxCCO. At the same time, the dual set of SDS also enables the use of multi-distance algorithms, such as spatially-resolved spectroscopy (SRS) and Dual Slope (DS), for estimating placental oxygen saturation ($PlO_2$). FSM V2 is the first example of a new generation of wearable NIRS devices, paving the way for smart and advanced monitoring of placental health at bedside. The FSM V2 was successfully validated on a bench-testing, metrological platform based on optical phantoms, as well as against a gold-standard, time-domain (TD) NIRS system, before moving to preliminary examples of clinical application and usability on pregnancy monitoring trials at University College London Hospital (UCLH), conducted as part of an *in vivo* proof-of-concept study.

## 1 Introduction

Solutions involving optical imaging and diffuse spectroscopy technologies, such as near infrared spectroscopy (NIRS) and diffuse optical tomography (DOT), represent nowadays promising clinical tools for bedside, real-time, physiological monitoring of the placenta non-invasively, in a cost-effective and compact manner [1–4]. Optical devices can indeed retrieve quantitative information on both placental oxygenation and haemodynamics by estimating continuous, dynamic changes in the concentrations of oxy- ($HbO_2$) and deoxyhaemoglobin (HHb), as well as on mitochondrial activity within the placenta via changes in the concentration of oxidised cytochrome-c-oxidase (oxCCO), deep within the abdominal tissue. For the inclusion of the latter biomarker, the use of a larger number of wavelengths for the spectral sampling (typically 6 or more) is required via a derivative technique known as either broadband NIRS (NIRS) or multi-wavelength NIRS [5, 6].

Furthermore, absolute placenta saturation ($PlO_2$) can also be retrieved by tissue oximetric measurements with optical technologies that employ the use of multiple sources-to-detector distances (SDS), enabling for powerful data analysis via multi-distance algorithms, such as spatially-resolved spectroscopy (SRS) [7] and Dual Slope (DS) [8]

However, several challenges in terms of measurement sensitivity and accuracy exists when using photonics technology for monitoring the placenta *in vivo*, in particular: (1) the depth at which the placenta lays within the abdominal cavity, ranging on average from 1 to 2.5 cm, with maximums of even 4-5 cm skin-to-top-of-placenta [9, 10]; and (2) the presence of a number of abdominal overlayers (on average from 1- to 2-cm thick), including skin, adipose tissue, muscle (rectum and uterus), as well as liquid layers such as amniotic fluid and liquor [9, 10]. All the above contribute to the overall attenuation the light transabdominal penetration, thus reducing the signal that can be retrieve from the placenta, and provide confounding factors within the data reconstruction.

In addition to such physical and optical constraints, the logistic of the clinical environment demands for any monitoring system aimed at bedside implementation to be as compact and embeddable as possible within hospital and neonatal intensive care units (NICU). Thus, recent efforts have been made towards making optical devices wearable and wireless, so that they can anatomically fit patient's abdomens and take as little disruptive space as possible within routine clinical procedures.

Given all the above-mentioned challenges, we hereby present the design, development, validation and first *in vivo* clinical applicability examples of a novel multi-wavelength NIRS system, called FetalSenseM V2 (FSM V2), that is capable of sampling at the required depth in the human abdomen and with adequate precision and sensitivity to the biomarkers of interest in the placenta, yet with full ergonomic, wearable and wireless integration within the clinical settings.

## 2 Materials and Methods

The FSM V2 is a fully wireless and wearable, multi-wavelength, NIRS device based on LED illumination at six NIR wavelengths (782, 800, 818, 830, 848 and 888 nm). The device represents the latest iteration in the development of a fully industrialised prototype of a LED-based NIRS device for bedside placental monitoring. Compared to previous versions [6, 11], ergonomics, compactness and downscaling have been pushed towards minimising the fingerprint of the system on both the patients and the clinical environment, while maintaining the same performances and even increasing the number of channels from two (3 and 5 cm for FSM V1 and V1.5) to three. Indeed, FSM V2 employs two sources located on the flexible wings of the sensor body (Figure 1a), and three photodiodes placed symmetrically in its centre at 1 cm from each other. The configuration thus provides two sets of three source-to-detector distances (SDS) at 3, 4 and 5 cm. This allows not only to estimate dynamic, relative changes in placental concentrations of $HbO_2$, HHb and oxCCO at up to 30 Hz, but with the dual set of SDS it also enables the use of both types of multi-distance algorithms, i.e., SRS and DS, for estimating $PlO_2$. Technical specifications and components of the FSM V2 device are reported in Table 1.
The device can be easily clipped tightly to the abdomen by use of specific adhesive consumable (Figure 1b). It then connects with the laptop and interfaces with its own control software (Figure 1c) via Bluetooth (yet USB connection is also possible, and it provide battery charging to the unit). Each device comes with its own, optically characterised solid phantom [12] for preoperational status check, allowing to assess the correct working functionality of the unit before any measurement session (Figure 1d).
Finally, data analysis and processing can be conducted from the raw data acquired with FSM V2 via a cloud-based, automated dashboard [13] that provides conversion of the dataset in SNIRF format [14], data cleaning and preprocessing, data quality check, motion artefacts correction, as well as estimates of all the biomarkers of interest (Figure 1e).

| **Specifications** | **FetalSenseM V2** |
|---|---|
| Sources | 2x Custom LED arrays (OMC-705 IDE 784-881.00 |
| Detectors | 3x Photodiodes (Texas Instruments OPT101) |
| PCB | ISP1807 with nRF52 Bluetooth |
| Accelerometer + gyroscope | LSM6DSL |
| Battery | 3.7V, 1000mAh, 15h estimated |
| Wavelengths (centroid, power, bandwidth) | - 784 nm ± 15 nm, 20 mW, FWHM <35 nm<br>- 800 nm ± 15 nm, 20 mW, FWHM <35 nm<br>- 818 nm ± 15 nm, 20 mW, FWHM <35 nm<br>- 835 nm ± 15 nm, 20 mW, FWHM <35 nm<br>- 851 nm ± 15 nm, 20 mW, FWHM <35 nm<br>- 881 nm ± 15 nm, 20 mW, FWHM <35 nm |
| SDS | 3, 4 and 5 cm |
| Channels | 6 (2 sources, 3 detectors) |
| Acquisition rate | 30Hz |

*Table 1. Technical specifications and main components of the FSM V2.*

## 3 Results and discussion

### *3.1 Digital, metrological and physiological validation*

Digital, metrological and physiological validation of the FSM V2 was performed using a bench-testing platform, named OptoCENTAL, developed by Giannoni *et al* [12]. Four distinct protocols, using both digital phantoms (based on anatomically aware Monte Carlo simulations of light transport through the abdomen) as well as combinations of solid and liquid phantoms to replicate optical, physiological and geometrical properties of the placenta and its abdominal overlayers, were employed. The results are summarised here, as reported in detail in Giannoni *et al* [12].

Firstly, the results of the Monte Carlo simulations indicated a theoretical depth sensitivity to the placenta of FSM V2 up to 70% for a minimum skin-to-top-of-placenta depth of 6.1 mm, compared to 83% for a Time-Domain (TD) NIRS system used as a benchmark, i.e., the MAESTROS II [15].
Then, basic instrument performances assessment on a set of solid phantoms reported: (1) signal-to-noise ratio (SNR) up to 116.015±44.47 at for the 3 cm SDS and a minimum of 13.721±5.38 at 5 cm for standard optical properties ($\mu_a$ = 0.11 $cm^{-1}$, $\mu'_s$ = 6.74 $cm^{-1}$, at 800 nm); (2) noise sensitivity measured by average coefficient of variation CV between $10^{-3}$ and $10^{-4}$ (same standard optical properties); (3) measurement linearity on average at 0.9312 for varying absorption from $\mu_a$ = 0.11 $cm^{-1}$ to $\mu_a$ = 0.81 $cm^{-1}$ at 800 nm; (4) stability over 1 hour of measurement, and repeatability over three consecutive days within ±10% changes with respect to the average signal value.
Afterwards, assessment of physiological accuracy in reconstructing dynamic changes in $HbO_2$, HHb and oxCCO, as well as sensitivity to crosstalk errors for CCO reconstruction was verified via the use of homogenous, blood liquid phantoms, oxygenated via $O_2$ and deoxygenated via either $N_2$ or yeast. The dynamic changes in concentrations of all the chromophores of interest retrieved by the FSM V2 show consistent trends in accordance with the expected physiological changes in the liquid phantom. Furthermore, accuracy in reconstructing $StO_2$ was evaluated by comparing the FSM V2, on the same phantom, with the MAESTROS II system. SRS with FSM V2 present underestimation errors in $PlO_2$ during full oxygenation (mean difference error of -11.40%, maximum difference error of -14.45%), whereas underestimation errors in $PlO_2$ are identified with DS with FSM V2 during both deoxygenation with $N_2$ (mean difference error of -11.73%, maximum difference error of -27.06%) and with yeast (mean difference error of -19.18%, maximum difference error of -35.46%).
Finally, depth sensitivity assessment of the FSM V2 was experimentally validated on a multilayered, solid and liquid phantom, replicating placental physiology on the bottom medium (with the same liquid blood phantom as above), plus the optical attenuation of the abdominal overlayers, using solid phantoms on the top. Again, dynamic changes in concentrations of all the chromophores of interest were reconstructed for FSM V2 at each SDS, in the liquid phantom beneath the overlayers at a depth of 1.5 cm.

### *3.2 Long-term repeatability and reliability*

Long-term repeatability and reliability assessment was added to the previous bench-testing validation by running routine preoperational status checks with the FSM V2 devices on its own service solid phantom (Figure 1d) during different days and over the course of three different periods of use, at 1, 3 and 6 months, respectively. The FSM V2 was tested in two different environmental settings to validate against operational variability. One was a laboratory setting, where measurements were almost daily, at the same time of the day, with the room lights off and without ever taking the unit off its phantom. The second setting was the routine clinical environment where the pilot studies on the device have been conducted at UCLH, with variable conditions every time and random time of use. Essentially, these were the prechecks performed by users before any measurement session on patients.
The results of both long-term assessment scenarios (Figure 2a) produced variability in the outputs within ±10% changes with respect to the average signal value, for all three period of use, thus corroborating a satisfactory reliability in the continuous use of the unit in any potential setting, from highly conservative to more chaotic.

### *3.3 Preliminary clinical application at UCLH*

Clinical, *in vivo* studies with FSM V2 are still ongoing at the time of writing at UCLH, UK and All India Institute of Medical Sciences (AIIMS), New Delhi, India.. Pilot results for clinical applicability of the device in the hospital setting are presented here (Figure 2b) to demonstrate preliminary capability of the FSM V2 in placental physiological monitoring at bedside, as well as in quantitative reconstruction of changes in $HbO_2$, HHb, oxCCO and $PlO_2$.

Across the two representative recordings, FSM V2 enabled continuous bedside monitoring over extended acquisition periods, with simultaneous reconstruction of haemodynamic and metabolic changes at multiple source–detector separations. The $HbO_2$, HHb and oxCCO traces showed temporally resolved fluctuations, including distinct transient events that were consistently detectable across measurement depths, demonstrating the system's sensitivity to dynamic changes in placental physiology. $PlO_2$ estimates were also continuously recovered using both SRS and dual-slope approaches, with the dual-slope measurements showing comparatively stable trends throughout the recordings, while the SRS estimates displayed greater patient- and channel-dependent variability. Overall, these preliminary measurements demonstrate the feasibility of obtaining continuous, depth-resolved haemodynamic, metabolic and oxygen-saturation information from the placenta in a clinical bedside environment.

## 4 Conclusion

The FSM V2 represents a further step towards the translation of diffuse optical technologies into continuous, non-invasive placental monitoring at the bedside. The system combines a compact, wearable and wireless architecture with six-wavelength illumination and multiple source–detector separations, enabling simultaneous assessment of changes in $HbO_2$, HHb and oxCCO, together with estimates of deep placental $PlO_2$. Bench testing demonstrated adequate sensitivity, stability and physiological tracking performance under controlled conditions, while the long-term assessment showed reproducible operation over up to six months of use, including within the more variable clinical environment. Preliminary measurements performed at UCLH further demonstrated the feasibility of acquiring continuous haemodynamic, metabolic and oxygenation-related information from the placenta during in vivo pregnancy monitoring. Although further clinical validation is required to establish measurement accuracy, repeatability and physiological interpretation across a larger and more diverse patient population, these initial results support the potential of FSM V2 as a practical platform for longitudinal assessment of placental function. Ongoing clinical studies will help determine its depth sensitivity to clinically relevant changes in placental physiology and its potential role within future antenatal monitoring pathways.

**Acknowledgement:** LG, UH, MT, DG, FL, NRZ, SM and IL were supported by the Wellcome Trust (219610/Z/19/Z), Wellcome Leap as part of the In Utero programme, Wellcome / EPSRC Centre for Interventional and Surgical Sciences (WEISS), and National Institute for Health Research University College London Hospitals Biomedical Research Centre. LG, FL, AA and IL are supported by: (1) the HyperProbe project, as part of the European Union's Horizon Europe research and innovation programme under grant agreement No 101071040 and of the UKRI grant number 10048387; (2) the fastMOT project, as part of the EU's HORIZON EUROPE programme under grant agreement number 101099291 and of the UK Research and Innovation (UKRI) under the UK government's Horizon Europe funding guarantee (grant number 10063660): and (3) the Medical Research Council (MRC) grant RESPECT4NEURODEVELOPMENT under grant agreement number EP/W035154/1.

**Conflict of Interests or disclosures:** The authors declare that there are no financial interests, commercial affiliations, or other potential conflicts of interest that could have influenced the objectivity of this research.

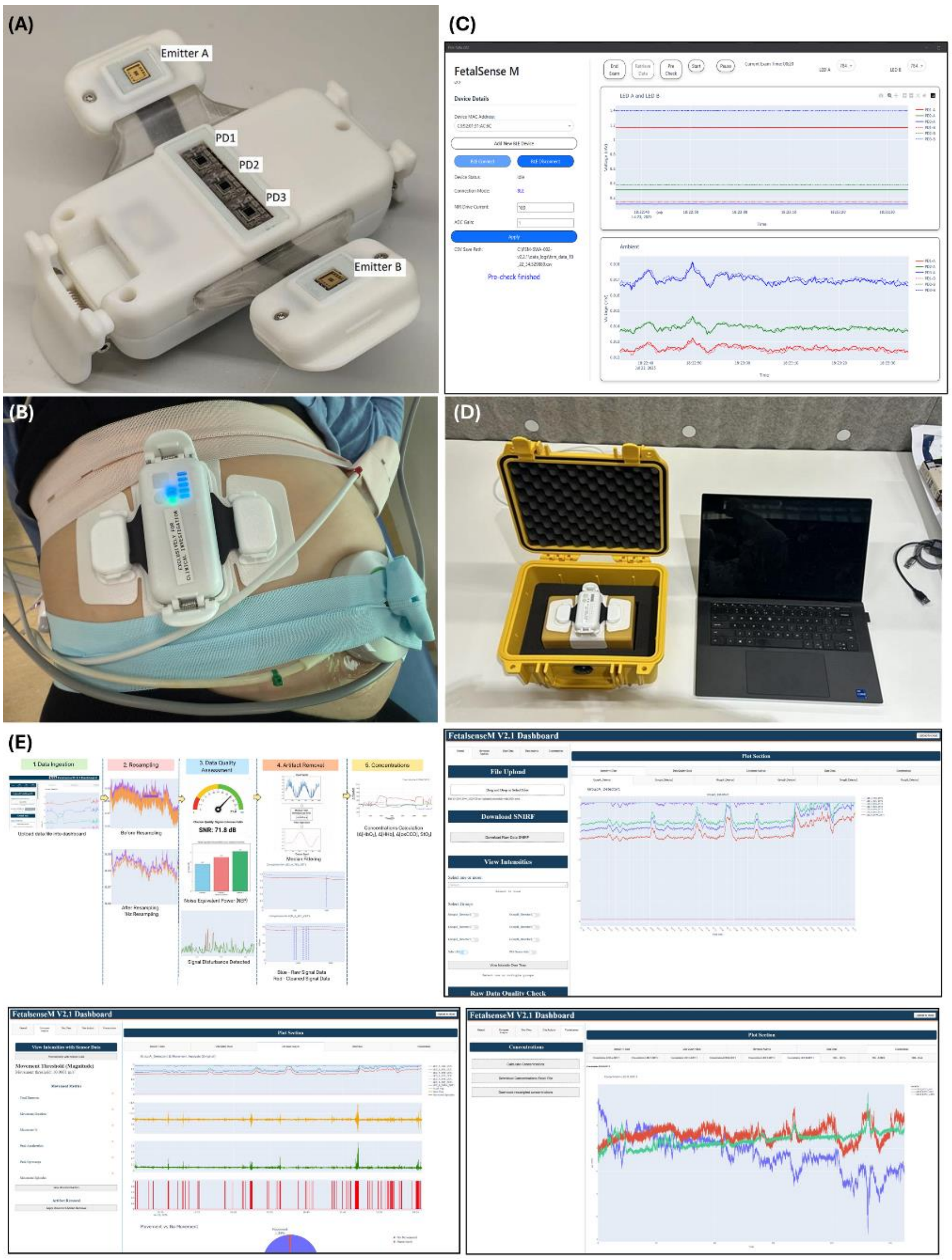

*Figure 1. (A) The FSM V2 device body with the wings and main body, labelling sources (LED A and B), as well as photodiode detectors (PD1, PD2 and PD3); (B) Example of the FSM V2 attached on a pregnant patient's abdomen; (C) The control software GUI of FSM V2; (D) The case of the FSM V2 with its own pre-check solid phantom and laptop; (E) The cloud-based, automated dashboard workflow (top-left) and examples of the interfaces used for the raw data processing (top-right), motion artefacts correction (bottom-left) and concentration estimates (bottom-right) of the FSM V2 outputs.*

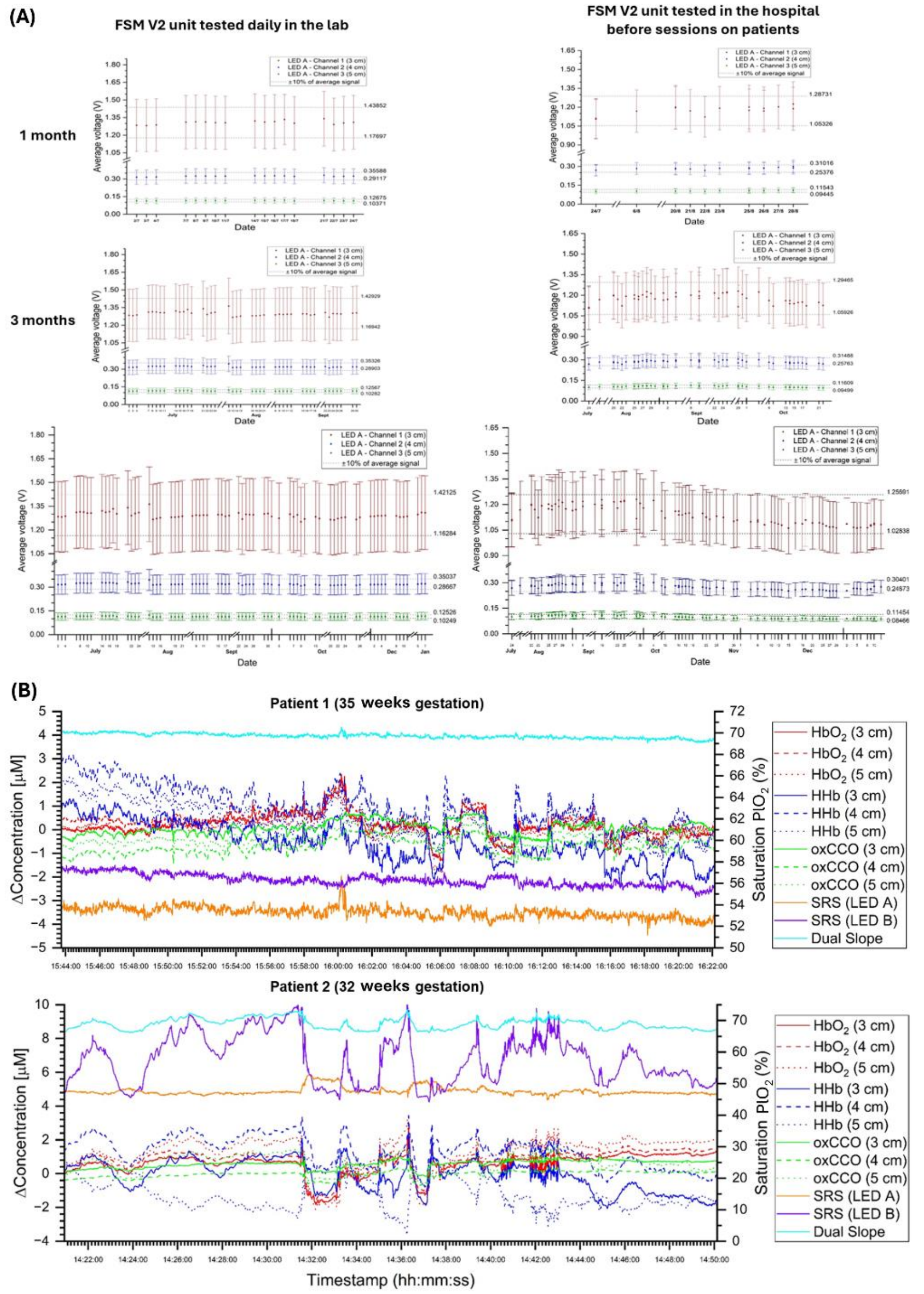

*Figure 2. (A) Results of the long-term reliability assessment of FSM V2, for 1, 2 and 6 months of use. Two case studies were compared, between prechecks run in a laboratory setting (higher repeatability, more conservative environment), as well as in routing clinical setting, before any measurement session on patients. All cases report variability in the outputs within ±10% changes with respect to the average signal value; (B) Examples of pilot data obtain during clinical application of the FSM V2 at UCLH on two different patient's pregnancy, with different gestational stages and different physiological tracking results in their session.*